# Confinement of antihydrogen for 1000 seconds

G.B. Andresen[1], M.D. Ashkezari[2], M. Baquero-Ruiz[3], W. Bertsche[4], E. Butler[5], C.L. Cesar[6], A. Deller[4], S. Eriksson[4], J. Fajans[3#], T. Friesen[7], M.C. Fujiwara[8,7], D.R. Gill[8], A. Gutierrez[9], J.S. Hangst[1], W.N. Hardy[9], R.S. Hayano[10], M.E. Hayden[2], A.J. Humphries[4], R. Hydomako[7], S. Jonsell[11], S. Kemp[5§], L. Kurchaninov[8], N. Madsen[4], S. Menary[12], P. Nolan[13], K. Olchanski[8], A. Olin[8&], P. Pusa[13], C.Ø. Rasmussen[1], F. Robicheaux[14], E. Sarid[15], D.M. Silveira[16], C. So[3], J.W. Storey[8$], R.I. Thompson[7], D.P. van der Werf[4], J.S. Wurtele[3#], Y. Yamazaki[16¶].

[1]Department of Physics and Astronomy, Aarhus University, DK-8000 Aarhus C, Denmark.

[2]Department of Physics, Simon Fraser University, Burnaby BC, V5A 1S6, Canada

[3]Department of Physics, University of California, Berkeley, CA 94720-7300, USA

[4]Department of Physics, Swansea University, Swansea SA2 8PP, United Kingdom

[5]Physics Department, CERN, CH-1211, Geneva 23, Switzerland

[6]Instituto de Fısica, Universidade Federal do Rio de Janeiro, Rio de Janeiro 21941-972, Brazil

[7]Department of Physics and Astronomy, University of Calgary, Calgary AB, T2N 1N4, Canada

[8]TRIUMF, 4004 Wesbrook Mall, Vancouver BC, V6T 2A3, Canada

[9]Department of Physics and Astronomy, University of British Columbia, Vancouver BC, V6T 1Z1, Canada

[10]Department of Physics, University of Tokyo, Tokyo 113-0033, Japan

[11] Department of Physics, Stockholm University, SE-10691, Stockholm, Sweden

[12]Department of Physics and Astronomy, York University, Toronto, ON, M3J 1P3, Canada

[13]Department of Physics, University of Liverpool, Liverpool L69 7ZE, United Kingdom

[14]Department of Physics, Auburn University, Auburn, AL 36849-5311, USA

[15]Department of Physics, NRCN-Nuclear Research Center Negev, Beer Sheva, IL-84190, Israel

[16]Atomic Physics Laboratory, RIKEN, Saitama 351-0198, Japan

[#]Also affiliated with Lawrence Berkeley National Laboratory, Berkeley, CA 94720, USA

[&]Also affiliated with Department of Physics and Astronomy, University of Victoria, Victoria BC, V8P 5C2 Canada

[§] Present address: Department of Physics, Durham University, Durham DH1 3LE, UK

[$]Present address: Physik-Institut, Zürich University, CH-8057 Zürich, Switzerland

[¶]Also affiliated with graduate School of Arts and Sciences, University of Tokyo, Tokyo 153-8902, Japan

(ALPHA Collaboration)

**Abstract**

**Atoms made of a particle and an antiparticle are unstable, usually surviving less than a microsecond. Antihydrogen, made entirely of antiparticles, is believed to be stable, and it is this longevity that holds the promise of precision studies of matter-antimatter symmetry. We have recently demonstrated trapping of antihydrogen atoms by releasing them after a confinement time of 172 ms. A critical question for future studies is: how long can anti-atoms be trapped? Here we report the observation of anti-atom confinement for 1000 s, extending our earlier results by nearly four orders of magnitude. Our calculations indicate that most of the trapped anti-atoms reach the ground state. Further, we report the first measurement of the energy distribution of trapped antihydrogen which, coupled with detailed comparisons with simulations, provides a key tool for the systematic investigation of trapping dynamics. These advances open up a range of experimental possibilities, including precision studies of CPT symmetry and cooling to temperatures where gravitational effects could become apparent.**

Experiments with atoms that do not exist naturally can be powerful tools for the study of fundamental physics [1,2, 3,4,5]. A major experimental challenge for such studies is the short intrinsic lifetimes of the exotic atoms. Atomic hydrogen is presumably stable [6] and, according to the CPT (Charge-Parity-Time reversal) theorem [7], antihydrogen - an atomic bound state of an antiproton and a positron [8,9] - should have the same lifetime. If sufficiently long confinement of antihydrogen can be achieved, a variety of possibilities will become available for fundamental studies with atomic antimatter. Examples include precision tests of CPT via laser [10] and microwave [11] spectroscopy on very few, or even a single trapped anti-atom; and laser [12,13,14] and adiabatic [15,16] cooling of antihydrogen to temperatures where gravitational effects become apparent.

In the first demonstration of antihydrogen trapping [17], the confinement time, defined by the time between the end of antihydrogen production and the shutdown of the magnetic trap, was set to 172 ms, the shortest time operationally possible. This maximized the chance of detecting rare occurrences of trapped antihydrogen before they could be lost. While a confinement time of a few 100 ms should be sufficient for initial attempts at antihydrogen spectroscopy [11], a critical question for future fundamental studies remains: what is the storage lifetime of trapped anti-atoms?

Reported trapping times of magnetically confined (matter) atoms range from <1 s in the first, room temperature, traps [18] to 10 to 30 minutes in cryogenic devices [19, 20,21,22]. However, antimatter atoms can annihilate on background gases. Also, the loading of our trap (*i.e.*, anti-atom production via merging of cold plasmas) is different from that of ordinary atom traps, and the loading dynamics could adversely affect the trapping and orbit dynamics. Mechanisms exist for temporary magnetic trapping of particles (*e.g.*, in quasi-stable trapping orbits [23], or in excited internal states [24]); such particles could be short-lived with a trapping time of a few 100 ms. Thus, it is not *a priori* obvious what trapping time should be expected for antihydrogen.

In this article, we report the first systematic investigations of the characteristics of trapped antihydrogen. These studies were made possible by significant advances in our trapping techniques subsequent to Ref. [17]. These developments, including incorporation of evaporative antiproton cooling [25] into our trapping operation, and optimisation of autoresonant mixing [26], resulted in up to a factor of five increase in the number of trapped atoms per attempt. A total sample of 309 trapped antihydrogen annihilation events was studied, a large increase from the previously published 38 events. Here we report trapping of antihydrogen for 1000 s, extending earlier results [17] by nearly four orders of magnitude. Further, we have exploited the temporal and spatial resolution of our detector system to perform a detailed analysis of the antihydrogen release process, from which we infer information on the trapped antihydrogen kinetic energy distribution.

The ALPHA antihydrogen trap [27,28] is comprised of the superposition of a Penning trap for antihydrogen production and a magnetic field configuration that has a three-dimensional minimum in magnitude (Fig. 1). For ground-state antihydrogen, our trap well-depth is 0.54 K (in temperature units). The large discrepancy in the energy scales between the magnetic trap depth (~50 μeV), and the characteristic energy scale of the trapped plasmas (a few eV) presents a formidable challenge to trapping neutral anti-atoms.

CERN's Antiproton Decelerator provides bunches of $3 \times 10^7$ antiprotons, of which $\sim 6 \times 10^4$ with energy less than 3 keV are dynamically trapped [29]. These keV antiprotons are cooled [30], typically with ~50% efficiency, by a preloaded cold electron plasma ($2 \times 10^7$ electrons), and the resulting plasma is radially compressed [31,32]. After electron removal and evaporative cooling [25], a cloud of $1.5 \times 10^4$ antiprotons at ~100K, with radius 0.4 mm and density $7 \times 10^7$ cm$^{-3}$ is prepared for mixing with positrons. Independently, the positron plasma, accumulated in a Surko-type buffer gas accumulator [33,34], is transferred to the mixing region, and is also radially compressed. The magnetic trap is then energized, and the positron plasma is cooled further via evaporation, resulting in a plasma with a radius of 0.8 mm and containing $1\times10^6$ positrons at a density of $5\times10^7$ cm$^{-3}$ and a temperature of ~40 K.

The antiprotons are merged with the positron plasma via autoresonant excitation of their longitudinal motion [26]. The self-regulating feature of this nonlinear process allows robust and efficient injection of antiprotons into the positrons with very low kinetic energies.

About $6\times10^3$ antihydrogen atoms are produced by allowing the plasmas to interact for 1 s. Most of the atoms annihilate on the trap walls [35], while a small fraction is trapped. A series of fast electric field pulses is then applied to clear any remaining charged particles. After a specified confinement time for each experimental cycle, the superconducting magnets for the magnetic trap are shut down with a 9 ms time constant. Antihydrogen, when released from the magnetic trap, annihilates on the Penning trap electrodes. The antiproton annihilation events are registered using a silicon vertex detector [36,37]. For most of the data presented here, an axial, static, electric bias field of 500 V·m$^{-1}$ was applied during the confinement and shutdown stages to deflect bare antiprotons which may have been trapped via the magnetic mirror effect [17]. This bias field ensured that annihilation events could be only produced by neutral antihydrogen.

The silicon vertex detector, surrounding the mixing trap in three layers (Fig. 1 a), allows position-sensitive detection of antihydrogen annihilations even in the presence of a large amount of scattering material (superconducting magnets and cryostat) [38], and is one of the unique features of ALPHA (*Methods*). The vertex detection and fast trap shutdown capabilities of our apparatus provide an increase in signal-to-background ratio against cosmic rays [39] by six orders of magnitude compared to the apparatus described in [40]. Improvements in annihilation event identification have also resulted in an increase in detection efficiency *(Methods)* relative to our previous work [17]. Knowledge of annihilation positions also provides sensitivity to the antihydrogen energy distribution, as we will show.

In Table 1 and Fig. 2, we present the results for a series of measurements, wherein the confinement time was varied from 0.4 s to 2000 s. These data, collected under similar conditions, contained 112 detected annihilation events out of 201 trapping attempts. Annihilation patterns in both time and position (Fig. 3) agree well with those predicted by simulation (see below). Our cosmic background rejection [39] allows us to establish, with high statistical significance, the observation of trapped antihydrogen after long confinement times (Fig. 2b). At 1000 s, the probability that the annihilation events observed are due to a statistical fluctuation in the cosmic ray background (*i.e.*, the Poisson p-value [6]) is less than $10^{-15}$, corresponding to a statistical significance of 8.0 σ. Even at 2000 s, we have an indication of antihydrogen survival with a p-value of $4\times10^{-3}$ or a statistical significance of 2.6 σ. The 1000 s observation constitutes a more than a 5000-fold increase in measured confinement time relative to the previously reported lower limit of 172 ms [17].

Possible mechanisms for antihydrogen loss from the trap include annihilations on background gas, heating via elastic collisions with background gas, and the loss of a quasi-trapped population [23] (see below). Spin-changing collisions between trapped atoms [22] are negligible because of the low antihydrogen density. The main background gases in our cryogenic vacuum are expected to be He and $H_2$. Our theoretical analysis of antihydrogen collisions indicates trap losses due to gas collisions give a lifetime in the range of ~300 to $10^5$ s, depending on the temperature of the gas (*Methods*). The observed confinement time scale of ~1000 s is consistent with these estimates. Note that trapping lifetimes of 10 to 30 minutes are reported for cryogenic magnetic traps for matter atoms, comparable to our observations, and that collisions with the background gas are cited as the likely dominant loss mechanism [20,21].

Precision laser and microwave spectroscopy will likely require ground-state anti-atoms, and hence estimation of the quantum state distribution of antihydrogen is of considerable importance [41,42,43,44]. In all previous work involving un-trapped atoms only highly excited states have been experimentally identified.

Antihydrogen atoms produced by the three-body process (involving two positrons and an antiproton) [45,46] are created in excited states. De-excitation to the ground state takes place via cascades involving radiative and collisional (*i.e.*, between the atom and a positron) processes. The slowest radiative cascade proceeds via circular states: $l = n - 1$, which allows us to estimate an upper limit for the cascade time. Our calculations, based on semi-classical quantization of energy and radiative rates, including the effect of blackbody radiation, show conservatively that more than 99% of trapped antihydrogen atoms will be in the ground state after 0.5 s (*Methods*). Therefore, our observed long trapping times of >> 1 s imply that most anti-atoms reach the ground state before being released, indicating that a sample of ground-state antihydrogen atoms has been obtained for the first time.

We now turn to considerations of the energy distribution and the orbital dynamics of trapped antihydrogen. Information on the energy distribution is essential in understanding the antihydrogen trapping mechanism. In addition, knowledge of the orbital dynamics will be important in the realisation of spectroscopic measurements, since the anti-atoms will typically overlap with the resonant radiation in only a small region of the trap volume.

Shown in Fig. 3a are experimental and theoretical plots of time (*t*) versus axial position (*z*) of the annihilations of released antihydrogen. Annihilation time *t* is measured from the start of trap shutdown. A simulation of 40,000 trapped antihydrogen atoms (see below) is compared with experimental data for 309 annihilation events detected by the silicon vertex detector. Figures 3 b-d show projections of these data onto the *t*- and *z*-axes. For detailed comparison with simulations, we select events with -200 mm < *z* < 200 mm, and *t* < 30 ms, taking into account the detector solid angle and the trap shutdown time. We also restrict the analysis to confinement times <1 s, since longer times are not modelled in the simulations, resulting in 273 annihilation events.

We developed a simulation-based theoretical model in order to investigate the trapping dynamics and the release process (*Methods*). Our simulations start with ground-state antihydrogen atoms with a distribution of initial kinetic energies *E*. The antihydrogen energy is an important input into the simulations, as it has been the subject of some controversy. Early experiments [47,48] as well as theoretical calculations [49,50] suggested that antihydrogen atoms were formed epithermally with kinetic energies that were substantially higher than the positron temperatures, implying that a vanishing fraction of antihydrogen produced in conventional plasma merging schemes [47,48] had trappable energies. (See [42] for an alternative interpretation of the data). In our standard simulation, we assume that antiprotons are thermalized in the positron plasma at a temperature of $T_{e+}$ = 54 K before antihydrogen formation takes place. This may be justified by the low kinetic energies of the antiprotons in our autoresonant mixing procedure [26]. Figure 4b shows the initial kinetic energy distribution *f(E)*, for simulated antihydrogen atoms that were trapped and then released to hit the trap walls. The main part of the distribution is characterized by a function $f \sim E^{1/2}$, *i.e.*, the tail of a three-dimensional Maxwell-Boltzmann distribution. The shape of this tail is independent of $T_{e+}$ as long as $E<<kT_{e+}$. The contribution of the positron plasma rotational energy to the total kinetic energy is negligible in the present case (*Methods*).

Atoms with kinetic energy *E* greater than the trap depth can be temporarily confined in a trap when *E* is shared between the degrees of freedom. These quasi-trapped orbits, well known for magnetically trapped neutrons [23], could be stable for many seconds. The population (~10%) above the 0.54 K trap depth in Fig. 4b represents the simulated atoms trapped in these quasi-confined orbits.

During the release of atoms from the trap, the hierarchy of the relevant time scales in our model results in notable consequences for the dynamics: $\tau_{mix} \gg \tau_{shut} > \tau_{axial} > \tau_{radial}$, where $\tau_{shut}$ ~9 ms is the e-folding time of the currents in the magnetic trap during shutdown, $\tau_{axial}$ ~ few ms is the characteristic period for the antihydrogen axial (along $\hat{z}$) motion, $\tau_{radial}$ ~ few 0.1 ms is that for radial (transverse to $\hat{z}$) motion, and $\tau_{mix} >$ ~1 s is the time scale for coupling between the axial and radial motions, as observed in the simulations.

Figure 4 shows characteristics of the simulated antihydrogen dynamics. Fig. 4a gives the mapping of the simulated annihilation time *t* to the initial energy *E*, the quantity of interest. The mapping includes the effect of adiabatic cooling, which is expected from the relation $\tau_{shut} > \tau_{axial}, \tau_{radial}$. It is instructive to analyze the axial and radial degrees of freedom separately, since $\tau_{mix} \gg \tau_{shut}$ (*i.e.* they are largely decoupled on the time scale of the shutdown). Figure 4c shows the initial axial ($E_{ax}$) and radial ($E_{rad}$) energies as a function of *t*. The fact that *t* is largely correlated with $E_{rad}$ instead of $E_{ax}$ can be understood as follows. During trap shutdown, both the mirror ($B_{mirror}$) and octupole ($B_{oct}$) fields decay with a time constant of $\tau_{shut}$ ~ 9 ms. While the axial well-depth $D_{ax}$ follows $B_{mirror}$, the radial depth $D_{rad}$ (proportional to $[B_{oct}^2 + B_{sol}^2]^{1/2} - B_{sol}$, where $B_{sol}$ = 1 T is the static solenoidal base field) decays as ~$B_{oct}^2$ with a time constant $\tau_{shut}/2$ ~ 4.5 ms for small $B_{oct}$ (Fig. 4a). This implies that antihydrogen atoms are released radially; i.e. the radial motion becomes unconfined before the axial motion, providing a direct relationship between $E_{rad}$ and $D_{rad}$. Note that the difference between the time constant for $D_{rad}$ ~ 4.5 ms (Fig. 4a dashed line) and that for $E_{rad}$ ~ 6 ms (blue dots in Fig. 4c) is due to adiabatic cooling.

While it is intuitive that there is some correlation between *t* and *E*, it is perhaps less obvious that annihilation position *z* should exhibit sensitivity to *E* (as seen in Fig. 3d). This sensitivity comes mostly from the correlation between $E_{ax}$ and *z*. Figure 4d shows such a correlation, which roughly maps the axial well-depth profile (Fig 1c). This is because only atoms with large enough $E_{ax}$ can climb up the axial potential hills before being released radially from the well. Thus, our simulations suggest that the *t*- and *z*-distributions have largely orthogonal sensitivities to $E_{rad}$ and $E_{ax}$, the implications of which will be discussed below.

We now compare the predicted *t*- and *z*-distributions with the data. The distributions from the standard simulation are shown as filled histograms in Fig. 3 b, d. We also show predicted *t*- and *z*-distributions from simulations with thermal antihydrogen distributions having temperatures of 100 mK and 10 mK, as well as for distributions involving only the quasi-trapped population. Our experimental data (points with error bars in Fig. 3) are in good agreement with the standard simulation, but are clearly incompatible with the lower energy distributions. Also, the data do not support a scenario wherein a large fraction of atoms are confined to quasi-trapped orbits (Figs. 3 b,d). Given that there are no fit parameters in our standard simulation, the agreement between simulated and experimental results evident in Fig. 3 supports the validity of the basic picture of antihydrogen dynamics presented above, which includes a thermal distribution (at the positron temperature) for the initial antihydrogen energy *f(E)* (Fig. 4b) and the calculated energy-time mapping including the effect of adiabatic cooling (Fig. 4a). The exponential tail of the simulated *t*-distribution (Fig. 3c) reflects the $E^{1/2}$ power law of a thermal distribution, and its agreement with the data implies that our model is valid down to very low energies. If antihydrogen production is indeed thermal, as suggested here, the trappable fraction would scale as ~$T_{e+}^{-3/2}$, pointing to the importance of reducing $T_{e+}$ for increasing trapping rates. (Note that colder $T_{e+}$ does not necessarily help if antihydrogen production is epithermal, as suggested in earlier work [47,48]).

The kinetic energies with which antihydrogen atoms collide with the trap walls ($E_f$) are predicted to be very small, significantly lower than the initial energies $E$, due to the adiabatic cooling and the conversion of kinetic to potential energy near the trap walls (see $f(E_f)$ in Fig. 4b). This suggests the possibility of realizing a very cold source of spin-polarized antihydrogen by slowly ramping down one of the confining magnets.

The orthogonal sensitivity of our experiment to $E_{ax}$ and $E_{rad}$ discussed above suggests the possibility of measuring a direction-dependent temperature distribution. Figure 3e illustrates this idea. The red histogram shows the z-distribution for trapped antihydrogen having an anisotropic energy distribution with $E_{ax}$ ~1 mK but $E_{rad}$ ~ 0.5 K. The predicted z-distribution is strongly peaked as compared to the standard simulation (filled histogram), because the low $E_{ax}$ atoms cannot climb the axial potential hill. Such anisotropic temperature distributions could be realized, *e.g.,* by cooling in one dimension, either via laser or adiabatic cooling [16]. Note that the predicted *t*-distributions are similar for the two energy distributions shown in Fig. 3e, hence the z resolution is needed to distinguish them. It is the position sensitive detection of atom losses, a distinctive feature of antimatter atom traps, that provides sensitivity to anisotropic energy distributions.

In this article, we have reported the first systematic studies of trapped antihydrogen. The findings can be summarized as follows: (1) We have demonstrated confinement of antihydrogen atoms for 1000 s. Our calculations show these atoms are very likely in the atomic ground state after ~1 s, providing the first indication that anti-atoms have been prepared in the ground state, as required for precision spectroscopy; (2) From the distributions of annihilations in time and position of the released anti-atoms, information on the kinetic energy distribution of the trapped antihydrogen was obtained for the first time. Our data are consistent with a model in which antihydrogen is produced from antiprotons thermalized in a positron plasma. Furthermore, from our detailed simulation studies, several features of trapping dynamics have been identified, including the possibility of measuring anisotropic energy distributions.

The implications of long-time confinement are very significant for future experiments with antimatter atoms. In antihydrogen spectroscopy, the total atomic excitation rate scales as $D_H \cdot (I_{rad})^n$ for an $n$-photon process, where $D_H$ is the density of anti-atoms and $I_{rad}$ the radiation intensity. The long confinement we observe dramatically reduces the required level of both $D_H$ and $I_{rad}$ because the anti-atoms can be interrogated for a longer time period $t_c$. Combined with the increased trapping rate, our observations amount to an increase in the figure of merit $D_H \cdot t_c$ by more than four orders of magnitude relative to our previously published work [17]. Similarly, for laser cooling, the cooling rate scales as $I_{rad}^{\,n}$. The long confinement time makes it plausible to envisage significant cooling even with existing (relatively weak) radiation sources [12,13] by cooling atoms for a long period of time (in the absence of a strong heating mechanism). Adiabatic cooling, possibly in one dimension [16], could further reduce antihydrogen temperatures to the sub mK range, where gravitational effects will be significant. The work reported here is a substantial step towards such fundamental studies with atomic antimatter.

**Methods**

[Detector and analysis]: The readout of the Si detector is initiated ("triggered") by coincidence signals from two or more particles hitting the inner-layer detector, which is segmented into 32 sub-modules. Once triggered, a field programmable gate array controlled acquisition system collects information on the amount of charge registered in each of 30,000 strips in the detector. This information is then recorded in high channel density (48 channel) flash analog-to-digital convertors and written on a disc at the rate of 500 events per second via a VME bus. The data recorded on the disc are processed off-line. The strips registering charges greater than a threshold value are considered to have been "hit". The hit thresholds were calculated for each strip, using the real data to correct for any fluctuations in the noise level. If more than one adjacent strip is hit, they are formed into a cluster of hits, and the average position, weighted by the charges, is used to determine the hit position. Each event produces an ensemble of hits. A pattern recognition algorithm identifies a "track", a helical trajectory of a charged particle in a magnetic field, and an annihilation position, "vertex", is determined from two or more tracks. The vertex reconstruction algorithm has been improved since our previous analyses [17,39] and the antiproton annihilation vertex identification efficiency has increased by 21% to give a detection efficiency of 57±6%. Our main detector background is due to cosmic rays, and they are discriminated via their event topology at a 99.5% rejection efficiency [39]. The remaining 0.5% constitutes a background for the annihilation detection at the rate of 47±2 mHz, or 1.4±1 x$10^{-3}$ counts within the 30 ms detection time window per trapping attempt. The event selection criteria have been determined without direct reference to the trapping data, in order to avoid experimenter bias [39].

[Trapping lifetime estimate]: We infer our background gas densities from the antiproton lifetime of order 15000 s measured under similar conditions. In the energy range 10-1000 K the antiproton-atom annihilation cross section can be approximated by the Langevin form

$$\sigma = \sqrt{\frac{2\alpha}{E_c} \frac{e^2}{4\pi\varepsilon_o}}$$

where $\alpha$ is the polarizability of the neutral particle, $E_c$ the collision energy, $e$ the elementary charge and $\varepsilon_0$ the permittivity in vacuum. Using this form, we obtain a density of order $5\times10^{10}$ m$^{-3}$ for He and H$_2$, independent of the gas temperature, or $7\times10^{-14}$ mbar for an ideal gas at 10 K. Note that no significant number of bare charged particles or ions should be present during the confinement.

It is difficult to quantify the temperature of background gases in our apparatus, as there is a direct vacuum path linking the cryogenic trapping region to the room-temperature components. Therefore, we report here a range of loss rates, corresponding to gas temperatures of 10 K to 100 K.

The antihydrogen-He elastic cross section was calculated using the adiabatic potential by Strasburger *et al.* [M1]. Annihilation, which mainly occurs via the formation of a metastable antiproton-He nucleus, was obtained using the methods in [M2] extended to higher energies. For antihydrogen-H$_2$ collisions no calculations exist in the relevant energy range, but at lower [M3] and higher [M4] energies, cross sections are of similar order of magnitude as those for antihydrogen-He scattering.

We obtain a lifetime against annihilation of ~$10^5$ s at a background gas temperature of 10 K and ~$10^4$ s at 100 K. Losses via heating through elastic collisions with the background gas dominate if the gas is warmer than about 100 K, which gives a heating rate of 0.002 K·s$^{-1}$, while for a colder background gas the rate drops to $5\times10^{-6}$ K·s$^{-1}$ at 10 K. Thus an antihydrogen atom heats by 0.5 K on a time scale between 300 and $10^5$ s.

[Cascade calculation]: To accurately calculate the radiative cascade of excited antihydrogen atoms, the quantum decay rates between all states in a strong field would need to be obtained [M5]. Reference [M5] calculated that a complete *lm* mixture for *n*=35 had 90% of the population in the ground state after approximately 25 ms. To be conservative, we consider the most pessimistic (slowest) possible decay path. The important trends are: states with higher principal quantum number decay more slowly and states with higher angular momentum decay more slowly. Therefore, we modelled the decay of a low-field-seeking, circular state in a 1 T field for *n* = 55. The binding energy of this state is shifted from the B=0 value of 52 K to approximately 9 K because of the magnetic field. This state is probably too weakly bound to survive the electric fields of our trap, thus we expect our trapped atoms to decay more quickly than this state. We solved for the quantum decay rates in a 1 T field; the quantum decay rates are somewhat higher than for B=0 because the energy level spacings are greater in a magnetic field. We numerically solved the coupled rate equations and found that more than 95% of the population was in the ground state by 300 ms and more than 99.5% of the population was in the ground state by 400 ms when we did not include black body radiation. We estimate that our atoms experience black body radiation with a temperature between 10 K to 100 K. When including the effects of black body radiation, we found that more than 99% of the population was in the ground state by 500 ms.

[Antihydrogen simulation]: We simulated the motion of the antihydrogen atoms in our trap through a direct numerical solution of the classical equations of motion for the atom in the ground state. The equations of motion were propagated using a fourth order, adaptive step size Runge-Kutta algorithm. The atoms were created uniformly where the positron density was non-zero and with an initial velocity chosen from a Maxwell-Boltzmann velocity distribution in three dimensions. We allowed the atoms to propagate through the trap for ~200 ms before the shutdown. The duration of the propagation is randomized by ±20 ms in order to avoid any spurious effects arising from a fixed time period. In the simulations reported here, the contribution of the positron plasma rotation energy to the antihydrogen kinetic energy is ignored, since the former is of order of 2 K at the plasma edge, compared to the assumed positron thermal energy of $T_{e+}$ = 54 K for our standard simulation (the fact that the effect of the rotation energy is negligible was checked in a separate simulation). This temperature is sufficiently close to the measured value of ~40 K; a small change in $T_{e+}$ does not affect the dynamics, as long as $T_{e+} \gg$ 0.54 K (our trap depth). The magnetic fields were modelled by accurately fitting the magnetic field from each of the mirror coils and the octupole coils separately. The measured current in the coils, and the calculated effects of the eddy currents in the Penning trap electrodes, were used to obtain the decay curves of the magnetic fields. To compute the force on the atoms, the gradient of the magnitude of the magnetic field was calculated using a symmetrical two-point finite difference; we ensured the accuracy of this step by checking energy conservation of the motion before the trap shutdown.

The adaptive step size algorithm can sometimes allow the atom to move a few mm or more during one step which can lead to inaccuracies in determining when and where the antihydrogen hits matter; we decreased the step size when the atom was near the wall so that a time step would not take the atom deeply into matter.

While the simulation results reported in this article were obtained by assuming ground state antihydrogen, we also checked whether the cascade cooling of Ref. [M6,M7] would affect the *t*- and *z*-distributions for antihydrogen annihilations upon release from the trap. For this study, the atoms were started in a *n*=30 state and we solved their motion including the random radiative decay. We found that more atoms were trapped (as seen in Refs. [M6,M7]), but did not find a discernable change (within our experimental accuracies) in the *t*- and *z*-distributions. This is because at the end of cascade, the shape of the energy distribution of the trapped antihydrogen is similar to the no cascade case, following the $E^{1/2}$ power law.

**Acknowledgements**

This work was supported in part by CNPq, FINEP/RENAFAE (Brazil), NSERC, NRC/TRIUMF, AIF, FQRNT (Canada), FNU (Denmark), ISF (Israel), MEXT (Japan), VR (Sweden), EPSRC, the Royal Society and the Leverhulme Trust (UK) and DOE, NSF (USA). We are grateful to the AD team for the delivery of a high-quality antiproton beam.

**Author contributions**

All authors contributed significantly to this work.

**Competing Financial Interests statement**

We declare that the authors have no competing interests as defined by Nature Publishing Group, or other interests that might be perceived to influence the results and/or discussion reported in this article.

[46] Fujiwara, M.C. et al. Temporally controlled modulation of antihydrogen production and the temperature scaling of antiproton-positron recombination, Phys. Rev. Lett. **101**, 053401 (2008).

[47] Gabrielse, G. et al. First measurement of the velocity of slow antihydrogen atoms. Phys. Rev. Lett. **93**, 073401 (2004).

[48] Madsen, N. et al. Spatial distribution of cold antihydrogen formation. Phys. Rev. Lett. **94**, 033403 (2005).

[49] Robicheaux, F. et al. Simulations of antihydrogen formation. Phys. Rev. A **70**, 022510 (2004).

[50] Hu. S.X. et al. Molecular-dynamics simulations of cold antihydrogen formation in strongly magnetized plasmas. Phys. Rev. Lett. **95**, 163402 (2005).

**METHODS REFERENCES**

[M1] Strasburger, K., Chojnacki H. & Skolowska A. Adiabatic potentials for the interaction of atomic antihydrogen with He and He$^+$, J. Phys. B: At. Mol. Opt. Phys. **38**, 3091-3105 (2005).

[M2] Armour, E.A.G., Todd, A.C., Jonsell, S., Liu, Y., Gregory, M.R. & Plummer, M. The interaction of antihydrogen with simple atoms and molecules. Nucl. Instrum. Meth. B **266**, 363-368 (2008).

[M3] Gregory, M.R. & Armour, E.A.G. Hydrogen molecule-antihydrogen scattering at very low energies. Nucl. Inst. Meth. B **266**, 374-378 (2008).

[M4] Cohen, J.S. Molecular effects on antiproton capture by $H_2$ and the states of $\bar{p}$-p formed. Phys. Rev. A **56**, 3583-3596 (1997).

[M5] Topçu, T. & Robicheaux, F. Radiative cascade of highly excited hydrogen atoms in strong magnetic fields. Phys. Rev. A **73**, 043405 (2006).

[M6] Taylor, C.L., Zhang, J. & Robicheaux, F. Cooling of Rydberg $\bar{H}$ during radiative cascade. J. Phys. B: At. Mol. Opt. Phys. **39**, 4945-4959 (2006).

[M7] Pohl, T., Sadeghpour, H. R., Nagata, Y. & Yamazaki, Y. Cooling by spontaneous decay of highly excited antihydrogen atoms in magnetic traps. Phys. Rev. Lett. **97**, 213001 (2006).
22

# FIGURE CAPTIONS

**Figure 1**: (a) A schematic view of the ALPHA trap. Radial and axial confinement of antihydrogen atoms is provided by an octupole magnet (not shown) and mirror magnets, respectively. Penning trap electrodes are held at ~9 K, and have an inner diameter of 44.5 mm. A three-layer silicon vertex detector surrounds the magnets and the cryostat. A 1 T base field is provided by an external solenoid (not shown). An antiproton beam is introduced from the right, while positrons from an accumulator are brought in from the left. (b) The magnetic field strength in the *y-z* plane (*z* is along the trap axis, with *z*=0 at the centre of the magnetic trap). Green dashed lines in this and other figures depict the location of the inner walls of the electrodes. (c) The axial field profile, with an effective trap length of ~270 mm.  (d) The field strength in the *x-y* plane.  (e) The field strength profile along the *x*-axis.

**Figure 2**: (a) Antihydrogen trapping rate (the number of trapped antihydrogen atoms per attempt), as a function of confinement time.  An antihydrogen detection efficiency of 0.57±0.06, derived from an independent calibration, is assumed. The error bars represent uncertainties from counting statistics only. Scatter within subsets of the data suggest the presence of a systematic uncertainty at the level of ±0.2 in trapping rate, which is not explicitly included; this does not affect our conclusions, nor our claims of statistical significance. (b) The statistical significance of observations against the cosmic ray background, in terms of  the number of standard deviations of the one-sided Gaussian distribution for each confinement time. The point for 0.4 s (>>20 sigma) is off scale, and is thus not shown.

**Figure 3**: (a) Time *t*- and axial *z*-distribution of annihilations upon release of antihydrogen from the magnetic trap for different confinement times (see legend), and comparison with simulation (grey dots). The simulation includes the effects of the annihilation detection *z* position resolution (~5 mm) and the detection efficiency as a function of *z*, both determined from a dedicated detector Monte Carlo study. (b) Comparison of the *t*-distributions between the data (error bars), and simulations of trapped antihydrogen with various initial energy distributions (histograms); see text. The blue filled histogram represents our standard simulation, while the yellow and purple histograms are for simulations with antihydrogen temperatures of 100 mK and 10 mK, respectively. The green histogram is a simulation for only the quasi-trapped atoms. The vertical error bars represent counting statistics, while the horizontal error bars represent the bin size of 2.5 ms. Our time resolution is << 1 ms. (c) Comparison of the data to the standard simulation, shown on a log scale. (d) Comparison of the annihilation position *z* between the data (error bars) and various simulations (histograms). The vertical error bars represent counting statistics while the horizontal ones represent the bin size of 25 mm. The same color code as used in (b) applies. (e) Predicted *z*-distribution for an anisotropic energy distribution with an axial energy of ~1 mK and a radial energy of ~0.5 K (red), compared to that of the standard (isotropic) energy distribution (blue filled histogram), suggesting the possibility of direction sensitive determination of antihydrogen energies (see text).

**Figure 4**: Predictions of the standard simulation characterizing the dynamics of trapped antihydrogen atoms and their release (see text). Energies are given in units of temperature. (a) Blue dots: A scatter plot of the initial antihydrogen kinetic energy $E$ versus the time $t$ at which the atom collides with the trap walls, providing the mapping between $t$ and $E$ in our measurements. Red lines: the time evolution of the axial trap well-depth $D_{ax}$ (solid) and of the radial well-depth $D_{rad}$ (dashed), which decay with time constants of ~9 ms and ~4.5 ms, respectively, after the release is initiated. (b) Filled histogram: Distribution of the initial kinetic energy of trapped antihydrogen. The vertical dashed line represents our trap depth of 0.54 K, above which the population is quasi-trapped. Green line: power law showing $E^{1/2}$ associated with the tail of a Maxwell-Boltzmann distribution. Red histogram: Distribution of antihydrogen kinetic energy $E_f$ at the time of annihilation on the trap walls. (c) Scatter plot showing the axial and radial components of the initial antihydrogen kinetic energies versus the annihilation time $t$. Blue dots show the radial energy $E_{rad} = 1/2 \cdot m_H \cdot (v_x^2 + v_y^2)$, and red dots show the axial energy $E_{ax} = 1/2 \cdot m_H \cdot v_z^2$, where $v_x$, $v_y$, $v_z$, are $\hat{x}$, $\hat{y}$ and $\hat{z}$ components of the velocity, respectively, and $m_H$ is the mass of the antihydrogen atom. (d) Axial and radial components of the initial antihydrogen kinetic energies versus annihilation position, $z$. The same colour scheme as (c) applies.

**TABLE**

Table 1: Summary of antihydrogen confinement time measurements.

| Confinement Time (s) | 0.4 | 10.4 | 50.4 | 180 | 600 | 1000 | 2000 |
|---|---|---|---|---|---|---|---|
| Number of attempts | 119 | 6 | 13 | 32 | 12 | 16 | 3 |
| Detected events | 76 | 6 | 4 | 14 | 4 | 7 | 1 |
| Estimated background | 0.17 | 0.01 | 0.02 | 0.05 | 0.02 | 0.02 | 0.004 |
| Statistical significance ($\sigma$) | >>20 | 8.0 | 5.7 | 11 | 5.8 | 8.0 | 2.6 |
| Trapped antihydrogen per attempt | 1.13 ±0.13 | 1.76 ±0.72 | 0.54 ±0.26 | 0.77 ±0.21 | 0.59 ±0.29 | 0.77 ±0.29 | 0.59 ±0.59 |

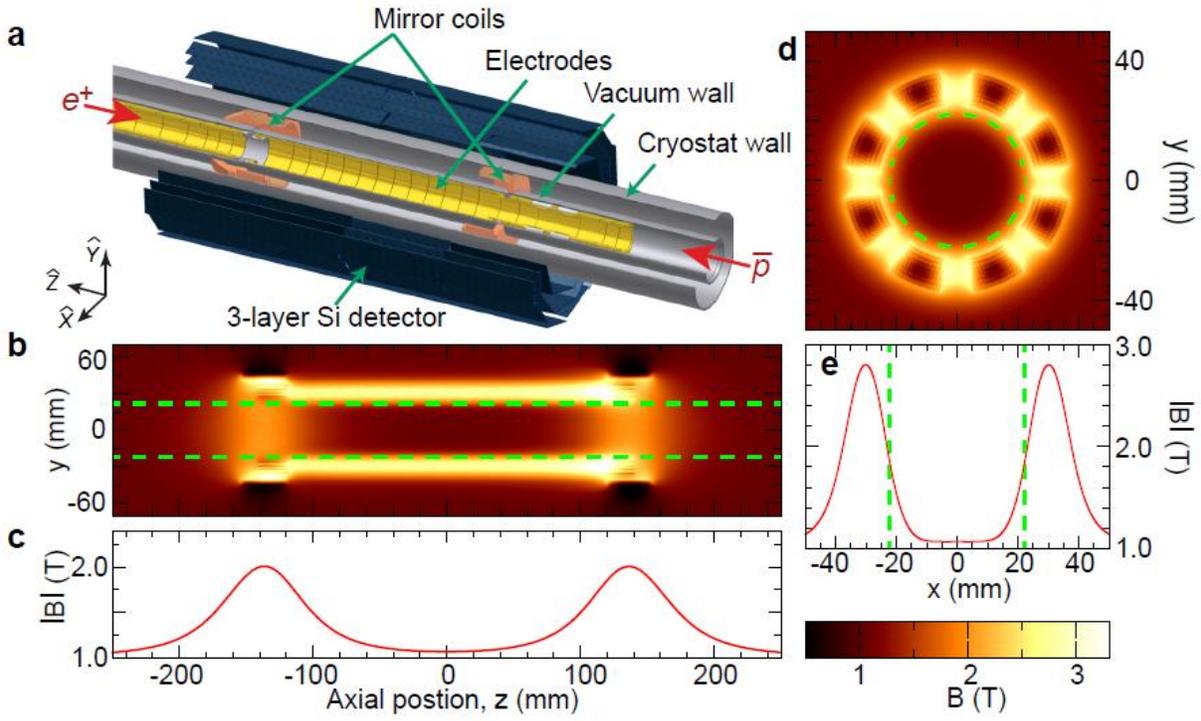

**Figure 1**

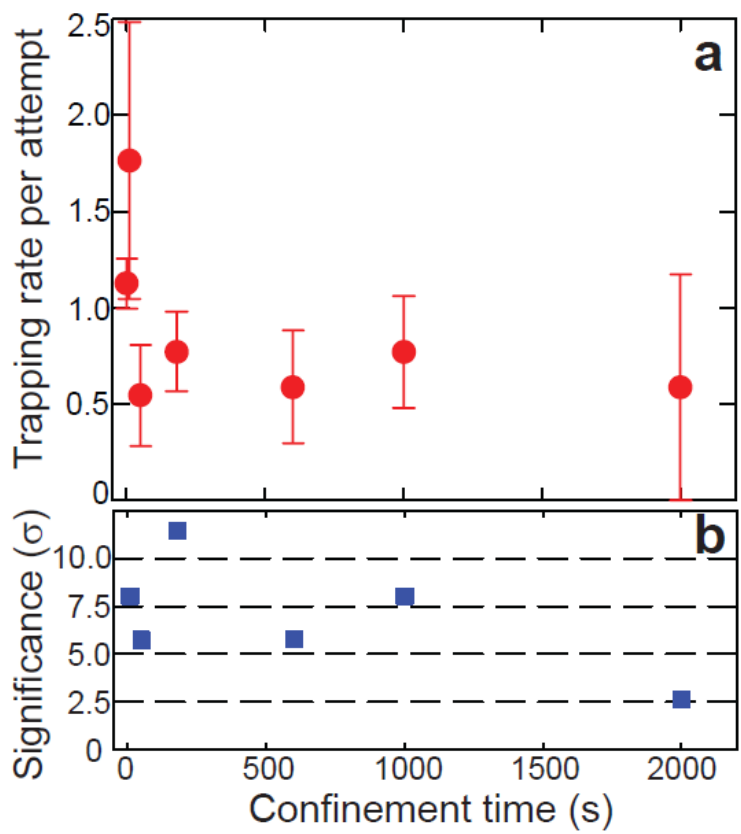

**Figure 2**

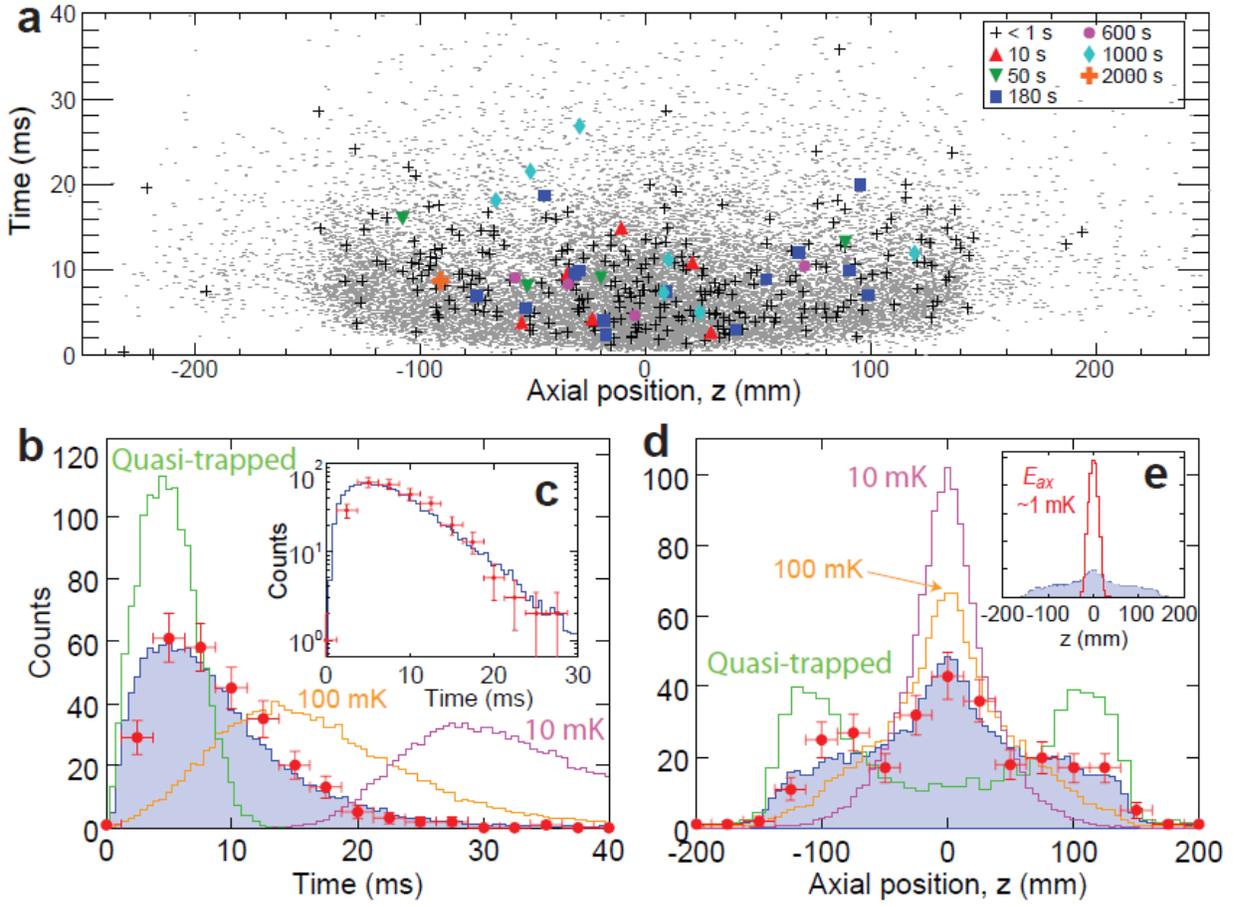

**Figure 3**

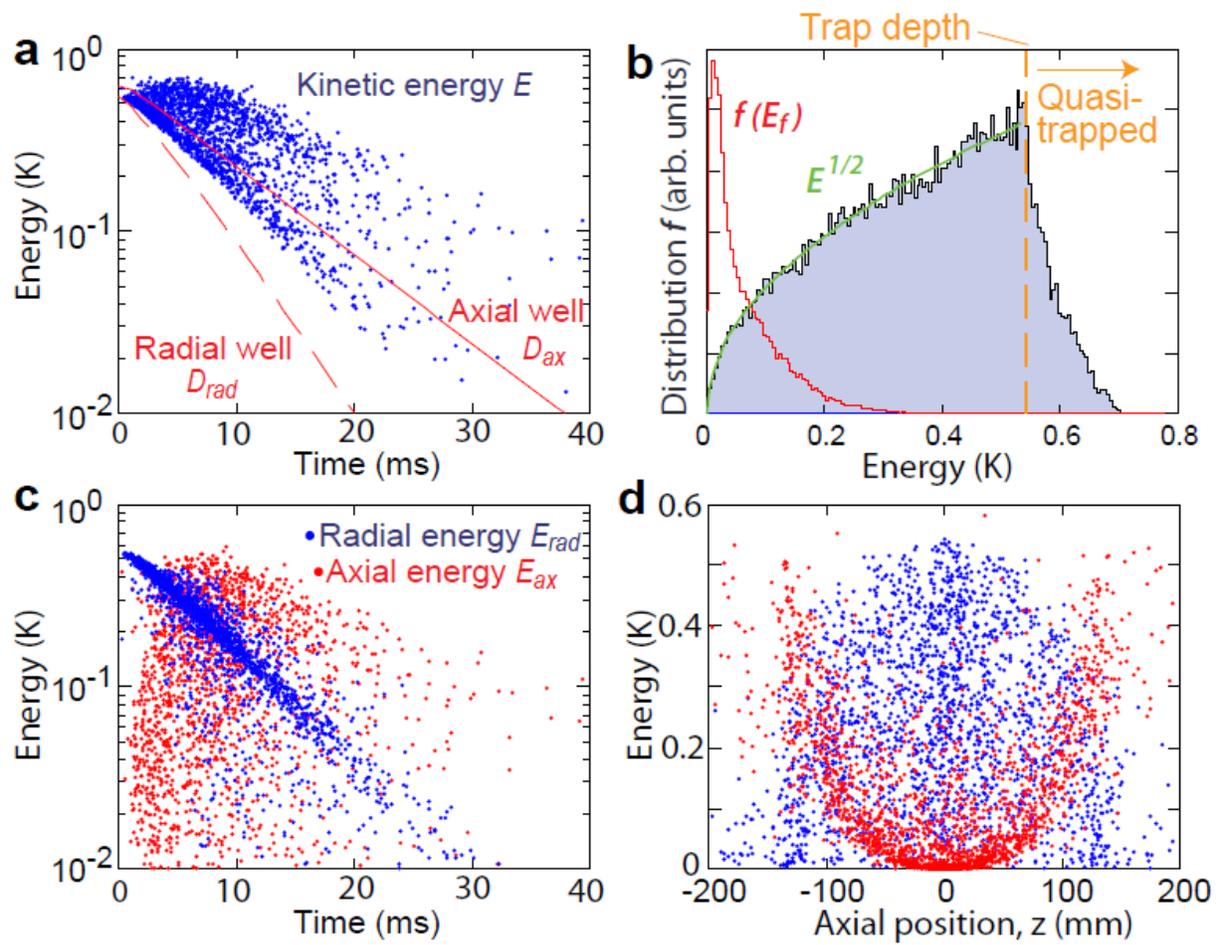

Figure 4